\documentclass{article}
\usepackage[utf8]{inputenc}
\usepackage{geometry}
\usepackage{amsmath}
\usepackage{graphicx}
\usepackage{amssymb}
\usepackage{braket}
\usepackage{chemformula}
\usepackage{setspace}
\usepackage{esint}
\usepackage{subscript}
\usepackage[lofdepth]{subfig}
\makeatletter

\usepackage{array}
\usepackage{babel}
\usepackage{authblk}
\usepackage[section]{placeins}
\usepackage{xcolor}
\usepackage{soul}
\usepackage{array, multirow, booktabs}
\usepackage{relsize}
\usepackage{ulem}
\usepackage{cite}
\usepackage{ctable}
\usepackage{longtable}
\usepackage{booktabs}
\usepackage[colorlinks=true, allcolors=blue]{hyperref}

\title{AtomVision: A machine vision library for atomistic images}
\author[1,2]{Kamal Choudhary}
\author[1]{Ramya Gurunathan}
\author[1]{Brian DeCost}
\author[3]{Adam Biacchi}

\affil[1]{Material Measurement Laboratory, National Institute of Standards and Technology, Gaithersburg, 20899, MD, USA}
\affil[2]{Theiss Research, La Jolla, 92037, CA, USA}
\affil[3]{Physical Measurement Laboratory, National Institute of Standards and Technology, Gaithersburg, 20899, MD, USA}

\begin{document}

\maketitle

\begin{abstract}
Computer vision techniques have immense potential for materials design applications. In this work, we introduce an integrated and general-purpose AtomVision library that can be used to generate, curate scanning tunneling microscopy (STM) and scanning transmission electron microscopy (STEM) datasets and apply machine learning techniques. To demonstrate the applicability of this library, we 1) generate and curate an atomistic image dataset of about 10000 materials, 2) develop and compare convolutional and graph neural network models to classify the Bravais lattices, 3) develop fully convolutional neural network using U-Net architecture to pixelwise classify atom vs background, 4) use generative adversarial network for super-resolution, 5) curate a natural language processing based image dataset using open-access arXiv dataset, and 6) integrate the computational framework with experimental microscopy tools. AtomVision library is available at \url{https://github.com/usnistgov/atomvision}.

\end{abstract}


\section{Introduction}







Only a few experimental techniques allow a materials scientist to ``see" the local atomic structure of a sample. Atomistic imaging techniques such as scanning tunneling microscopy (STM), atomic force microscopy (AFM), transmission electron microscopy (TEM) and their variants provide insights into the local atomic structure, defects and their dynamics, which are critically linked to the functionality and performance of the materials \cite{goodhew2000electron}. Due to rapid growth in computer-vision techniques \cite{krizhevsky2017imagenet,taigman2014deepface}, its application to atomic scale image data is natural. These data can be obtained from experimental as well as computational methods and recently their usage has become widespread\cite{holm2020overview,choudhary2022recent,vasudevan2019materials,modarres2017neural,ge2020deep,larmuseau2020compact,yang2018microstructural,hsu2021microstructure,chun2020deep,dai2021graph,stan2020optimizing,cohn2021instance,taillon2021nexuslims,decost2017uhcsdb,jesse2016big,schroff2010harvesting,hua2015prajna}. Nevertheless, an integrated library to capture, curate, generate datasets and apply data-analytics methods is still needed. 

Such libraries can be useful for microscopy image tasks such as image classification \cite{modarres2017neural,ziletti2018insightful,choudhary2021computational,schwenker2021exsclaim}, pixelwise learning (e.g., semantic segmentation)\cite{decost2017uhcsdb,madsen2018deep,maksov2019deep,yang2021deep,roberts2019deep,vlcek2019learning,ziatdinov2017learning,ovchinnikov2020detection}, object/entity recognition, localization, super-resolution \cite{de2019resolution,ede2020partial,rashidi2018autonomous,eppel2020computer,scime2020layer} etc. The application of such libraries encompasses multiple science domains such as materials science, condensed matter physics, biology etc \cite{agrawal2019deep,choudhary2022recent,ge2020deep}.

Computationally, there are several methods for simulating STM and scanning transmission electron microscopy (STEM) images. STM images can be computational simulated using Bardeen\cite{bardeen1961tunnelling}, Tersoff-Hamman\cite{tersoff1983theory} and Chen\cite{chen1994introduction} methods. STM images can be either constant height or constant current based. For standard high angle annular dark field (HAADF) STEM, methods such as convolution, Bloch wave and multislice approximations \cite{combs2019fast,kirkland1998advanced,kirkland2016computation,allen2003lattice,allen2003lattice,cowley1957fourier} can be used. The convolution approximation is one of the fastest ways to simulate STEM images. It is based on an incoherent linear image model that convolves the probe point-spread function with simple atomic potentials for the specimen and are usually used for thin films. Bloch wave and multislice methods are computationally heavy but are more generalizable.

Some of the major libraries (experimental and computational) based on microscopy images include  ab initio Transmission Electron Microscopy (abTEM) \cite{madsen2021abtem}, EXtraction, Separation, and Caption-based natural Language Annotation of IMages (EXCLAIM) \cite{schwenker2021exsclaim}, AtomAI\cite{ziatdinov2021atomai},  \cite{somnath2017pycroscopy}, Prismatic \cite{ophus2017fast} and Quantitative TEM/STEM Simulations (QSTEM)\cite{koch2002determination}. Deep learning techniques such as convolutional neural network (CNN) are commonly applied to atomistic image data. There have been several previous works on the application of deep-learning (DL) techniques \cite{choudhary2022recent} to atomistic image data. In Ref \cite{madsen2018deep} pixelwise DL was applied to detect atoms in simulated atomic-resolution TEM images of graphene. A neural network model was developed to detect the presence of atoms as well as predict its height. Ref \cite{maksov2019deep} demonstrated atomistic defect recognition and tracking across sequences of atomic-resolution STEM images of WS$_2$. In ref \cite{yang2021deep} U-net architecture was used to detect vacancies and dopants in WSe$_2$ in STEM images with high model accuracy. In ref \cite{roberts2019deep}  DefectSegNet was developed to automatically identify defects in transmission and STEM images of steel including dislocations, precipitates, and voids. Ziatdinov \textit{et al.} applied DL techniques to learn surface molecular structures  \cite{ziatdinov2017learning}. More details about previous works on atomistic imaging and machine learning techniques can be found elsewhere \cite{agrawal2019deep,choudhary2022recent,ge2020deep}.

In this work, we present the AtomVision library, which can be used to generate a simulated STM/STEM dataset using several levels of approximation, as well as using natural language processing to collect images from literature and experiments. We also provide generalized scripts that can be used for a broad level of image machine learning tasks such as identifying five 2D-Bravais lattices \cite{moeck2018accurate}, atomistic segmentation to distinguish between background and images and later apply convolutional neural network (CNN)/graph neural network (GNN) on the segmented images, and generative design of atomistic images using generative adversial network (GAN) techniques.  This library is a part of the NIST-JARVIS (Joint Automated Repository for Various Integrated Simulations) infrastructure \cite{choudhary2020joint} for accelerated materials design using electronic structure, force-field, machine learning calculations and experiments. The atomvision library is publicly available at \url{https://github.com/usnistgov/atomvision}.






\section{Methods}

\subsection{Dataset generation and curation}
We use Tersoff-Hamann (TH) technique \cite{tersoff1983theory} to calculate the STM images of 2D materials. TH is a simple model of an s-wave STM tip.

\begin{equation}
    n(r,E)=\sum_{\mu}{|\psi(r)^2|\delta(\epsilon_{\mu}-E)}
\end{equation}

\begin{equation}
\int_{E_F}^{E_F+eV} n(r,E) \,dE 
\end{equation}

In this approach, the tunneling current $I$, which depends on the tip position $r$ and the applied voltage $V$, is proportional to the integrated local density of states (ILDOS). The ILDOS is calculated from the Kohn-sham eigenvectors, $\psi_{\mu}$, and eigenvalues, $\epsilon_{\mu}$, where $\mu$ labels different states. $E_F$ is the Fermi-energy. Different experiments will choose different applied voltages, but we concentrate on two values, 0.5 eV for positive bias and -0.5 eV for negative bias, which require integrating from $E_F$ to $E_F\pm$0.5eV. We choose 0.5 eV range for simplicity sake, and other values usually produce qualitatively similar images for metals or small gap semiconductors. However, simulations for other voltages should also be possible with the method and tools discussed in this work. This method is readily available in DFT software such as Vienna ab initio simulation package (VASP) \cite{kresse1996efficiency,kresse1996efficient}.

The STEM images \cite{combs2019fast,kirkland1998advanced,kirkland2016computation,allen2003lattice,allen2003lattice,cowley1957fourier} were generated with the convolution approximation (based on fast Fourier transform based convolutions) following:

\begin{equation}
    I(r)=R(r,Z) \otimes PSF(r)
\end{equation}

where r is a 2D vector in the image plane, I(r) is the image intensity,
\begin{equation}
    R(r,Z) = \sum_{i}^{N} Z_{i}^{1.7}\delta(r-r_i)
\end{equation}

$R(r,Z)$ is the transmission function of the $N$ atoms at position $r_i$, and includes information about the atomic potential of the system given by $Z_i$, the atomic number of the atom. Rutherford scattering from the nuclear charge predicts a $Z^2$ dependence of the intensity, but the exponent is reduced by core electron screening, and depends on the detection collection angles. The power value of 1.7 is an approximate value that represents a compromise between these many factors. In previous works, the power values of 1.3-1.7 \cite{yamashita2018atomic} have also been used to match experiments, however, for the sake of generality, we use 1.7 for all the systems.  We note that for such STEM images only crystallographic coordinates and atom-type information are needed. The optimized geometry for the 2D materials were obtained from density functional theory calculations. We generate STEM images with output size of 256x256 pixels for at least 2.5 nm x 2.5 nm size. The microscope point spread function (PSF) is modeled as a normalized Gaussian with width of 0.5 $\AA$. We use the 2D materials available in the JARVIS-DFT-2D \cite{choudhary2020joint,choudhary2017high}, Computational 2D Materials Database (C2DB) \cite{haastrup2018computational} and 2DMatPedia \cite{zhou20192dmatpedia} datasets, leading to 9150 systems with unique chemical compositions and structural spacegroup information.  
 

For natural language processing (NLP) related dataset, we use the open access arXiv dataset, which consists of 1.8 million articles starting from 1986 to 2020. We use ChemNLP \cite{choudhary2022chemnlp} to extract chemistry information from the arXiv articles. We search for keywords such as STEM, STM, microscopy, HRTEM, scanning tunneling microsopy and scanning transmission electron microscopy in the abstract and figure captions of articles, and if the system has that info, we further find out if the article contains clear STM/STEM images to curate an image dataset. The figure caption parsing was carried out with the BeautifulSoup package.

Currently, the experimental STEM image dataset consists of images of nanoparticles. As an example experimental dataset we use TEM images of Iron oxide (Fe$_3$O$_4$), rhodium (Rh), and tin(II) sulfide (SnS) nanoparticles. In future, we plan to expand this dataset to multiple systems, especially for the materials available in the JARVIS-DFT dataset. 
Iron oxide, rhodium, and tin(II) sulfide nanostructures were synthesized using previously reported solution strategies based on the thermal decomposition of elemental precursors.  All syntheses were carried out under Ar using standard Schlenk techniques.  Briefly, Fe$_3$O$_4$ spherical nanoparticles were synthesized by heating iron oleate in benzyl ether to 300 °C and then centrifugally washing twice before dispersing the collected product in hexanes \cite{biacchi2022polyoxovanadates}.  Rh triangular nanoplates were synthesized by heating RhCl$_3$.xH$_2$O and 40,000 molecular weight poly(vinylpyrrolidone) in triethylene glycol to 135 °C before then centrifugally washing twice and redispersing the product in ethanol \cite{biacchi2015ligand}. $\alpha$-SnS micron-sized nanoribbons were synthesized by heating SnCl$_2$ and sulfur powder in oleylamine to 180 °C and then centrifullally washing twice before redispersing the product in toluene \cite{biacchi2018contact}.  Transmission electron microscopy (TEM) images were collected using a Phillips EM-400 operating at an accelerating voltage of 120 kV and high-resolution TEM (HRTEM) images were obtained with a FEI Titan 80-300 operating at 300 kV.  Samples were prepared by casting one drop of dilute nanomaterial solution onto 300-mesh Formvar and carbon-coated copper grids (Ted Pella). Please note that commercial products used in this work are identified to specify procedures. Such identification does not imply recommendation by National Institute of Standards and Technology (NIST).

\subsection{Machine learning model}

For the machine learning models, we primarily use the STEM dataset developed with the convolution approximation. We use several machine/deep learning approaches such as clustering, classification with convolution and graph convolution neural network, fully convolutional neural network using U-Net and generative adversial network \cite{holm2020overview,choudhary2022recent,vasudevan2019materials,modarres2017neural,ge2020deep,larmuseau2020compact,yang2018microstructural,hsu2021microstructure,chun2020deep,dai2021graph,stan2020optimizing,cohn2021instance,taillon2021nexuslims,decost2017uhcsdb,jesse2016big,schroff2010harvesting,hua2015prajna}.

For clustering analysis, we use t-distributed stochastic neighbor embedding (t-SNE), which is a statistical method for visualizing high-dimensional data in a two- or three-dimensional map. The t-SNE plot was generated with the help of Scikit-learn library \cite{pedregosa2011scikit}. The images were flattened into a python numpy array and then their dimensionality was reduced using nonlinear t-SNE for visualization purposes.

The pixelwise classification/semantic segmentation task was performed with segmentation-models-with-pytorch (SMP) \cite{Iakubovskii2019} package using U-Net \cite{ronneberger2015u} pretrained model using Binary cross-entropy with logits loss (BCELogitLoss) function. All the supervised ML tasks used 75:25 training:testing of samples during training.

For the Bravais lattice classification task, we use DenseNet (Dense Convolutional Network) \cite{huang2017densely} with pre-trained model available in PyTorch \cite{paszke2019pytorch}. We use a uniform size of 256x256 images for each material. We use Pytorch-Ignite library to setup the training run with Adam optimizer, 0.001 learning rate and negative log likelihood loss (NLLLoss) for 100 epochs.

After pixelwise classification, we convert the images into networkx and deep graph library graphs \cite{wang2019deep}, which are then used along with atomistic line graph neural network (ALIGNN) \cite{choudhary2021atomistic} for image classification tasks as well. We use maximum, minimum and mean intensity of blobs in the images as the node features, while a 4 $\AA$ cutoff is used to generate neighborlist and generate bond-angles of different nodes. We use a batch size of 32, learning rate of 0.001, AdamW optimizer, negative log likelihood loss (NLLLoss) and 50 epochs for ALIGNN training. We used the original hyperparameters of the ALIGNN model as used in ref.\cite{choudhary2021atomistic}.

We create a synthetic dataset of STEM images with low resolution (4 times lower resolution, i.e. 64x64 pixel images instead of 256x256) and high-resolution (as generated with convolution approximation) images and train a generative adversarial network (GAN) for image super-resolution (SR) using SR-GAN model \cite{ledig2017photo}. In SR-GAN, we use the 4th layer of VGG19 (visual geometry group  convolutional neural network that is 19 layers deep) \cite{simonyan2014very} as feature-extractor. We use a perpetual loss function during SR-GAN training, which is a combination of both adversarial loss and content loss. We train the model for 50 epochs, learning rate of 0.00008 and Adam optimizer during training.

Next, we analyze the reconstruction of image capabilities using an autoencoder model with PyTorch. We take the 256x256 image, and use an auto-encoder of dimension 1120. The decoding part of the model reconstructs the image in 256x256 size. We train the models for 200 epochs, with mean squared loss function, Adam optimizer, and a learning rate of 0.001. 


\section{Results}

\begin{figure}[hbt!]
    \centering
    \includegraphics[trim={0. 0cm 0 0cm},clip,width=0.98\textwidth]{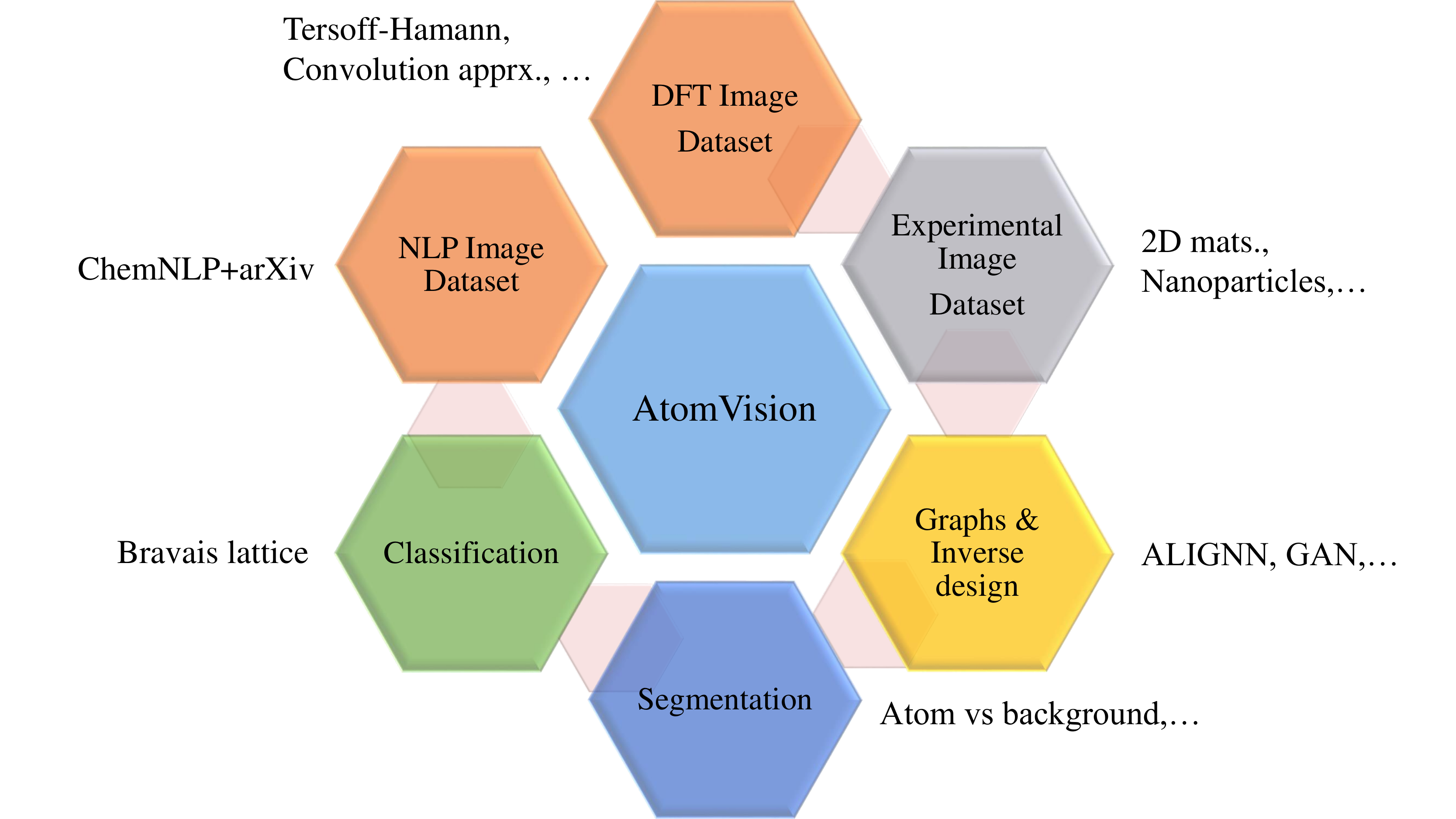}
    \caption{\label{fig:schematic}\textcolor{black}{A Schematic overview of the AtomVision library. AtomVision presents an integrated library of dataset and AI/ML tools for atomistic images. The tool can be used for generating and curating microscopy dataset in a systematic manner as well as apply machine learning tools on the image dataset.}}
\end{figure}

A schematic overview for AtomVision library is shown in Fig.~\ref{fig:schematic}. Usually any image analytic technique application would require a large dataset. In AtomVision, the dataset can be obtained from density functional theory, convolution approximation, natural language approximation, and experiments. For instance, STM images for 2D materials in both positive and negative biases were obtained from the Tersoff-Hamann approach as implemented in the JARVIS-Tools. The STM image consists of 1400 images for 2D materials in the JARVIS-DFT dataset. Although the application has been carried out for 2D materials, it can be applied to other non-2D systems as well. The STEM image dataset for 2D materials were obtained with the convolution approximation for systems in JARVIS-DFT, C2DB and 2DMatPedia computational datasets. While STM techniques require charge densities and wavefunctions to obtain integrated DOS values, the STEM dataset using the convolution approximation can be directly obtained with atomic types and coordinates information only.

\begin{figure}[hbt!]
    \centering
    \includegraphics[trim={0. 0cm 0 0cm},clip,width=0.98\textwidth]{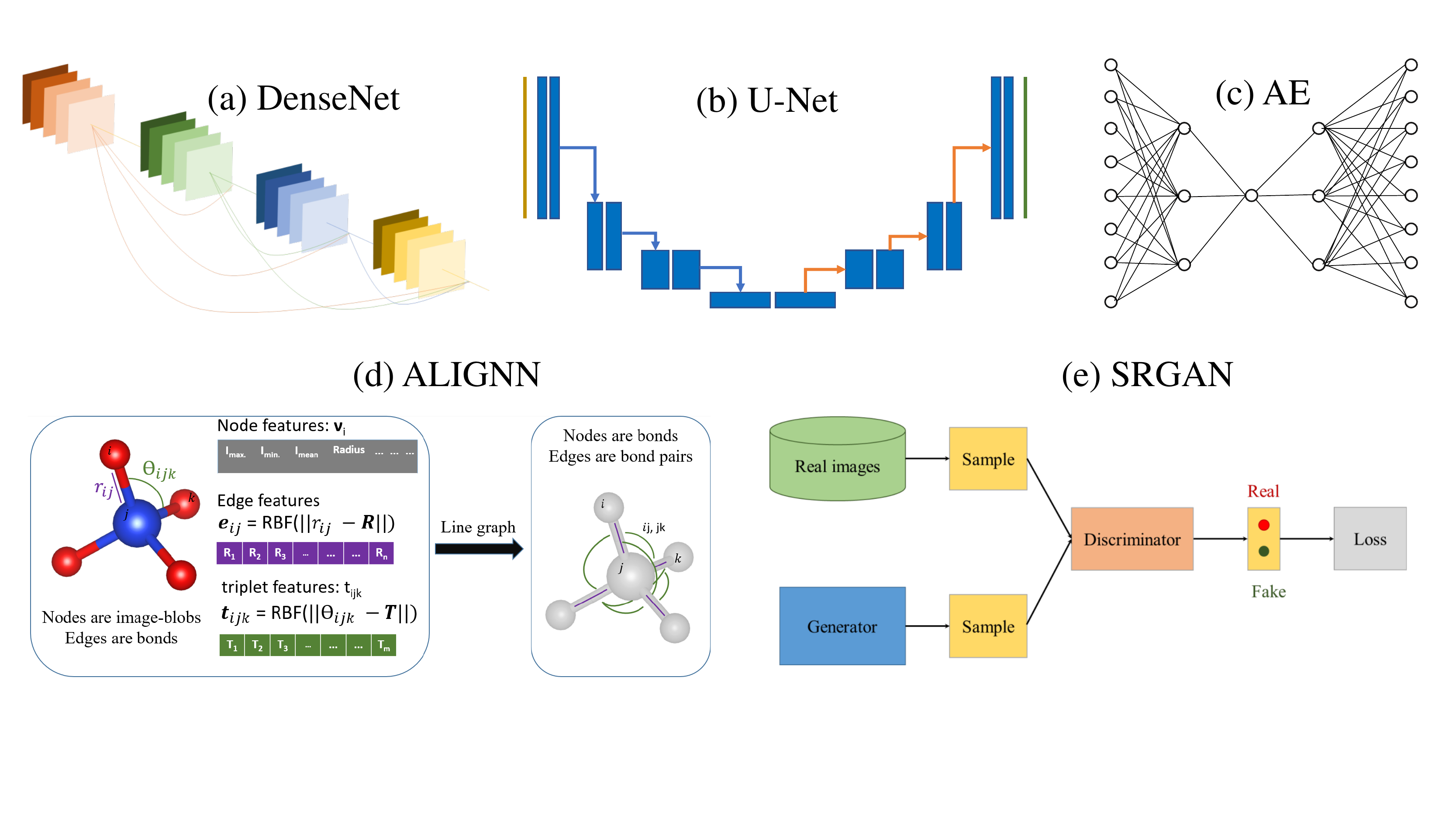}
    \caption{\label{fig:archs}\textcolor{black}{A few example machine learning architectures for images available in the AtomVision library. a) DenseNet is based on convolutional neural networks, b) U-net is a fully convolutional neural network, c) autoencoder, d) atomistic line graph neural network, e) super-resolution generative adversarial neural network. }}
\end{figure}

We also show some of the machine learning architectures adopted in AtomVision in Fig.~\ref{fig:archs}. These models are based on well-known deep learning models such as convolution neural network (CNN),  graph neural network (GNN), and generative models such as auto-encoders (AE) and generative adversarial networks (GAN). The applications of these architectures and their performances are discussed in detail below.

\subsection{Validation of dataset}
\begin{figure}[hbt!]
    \centering
    \includegraphics[trim={0. 0cm 0 0cm},clip,width=0.98\textwidth]{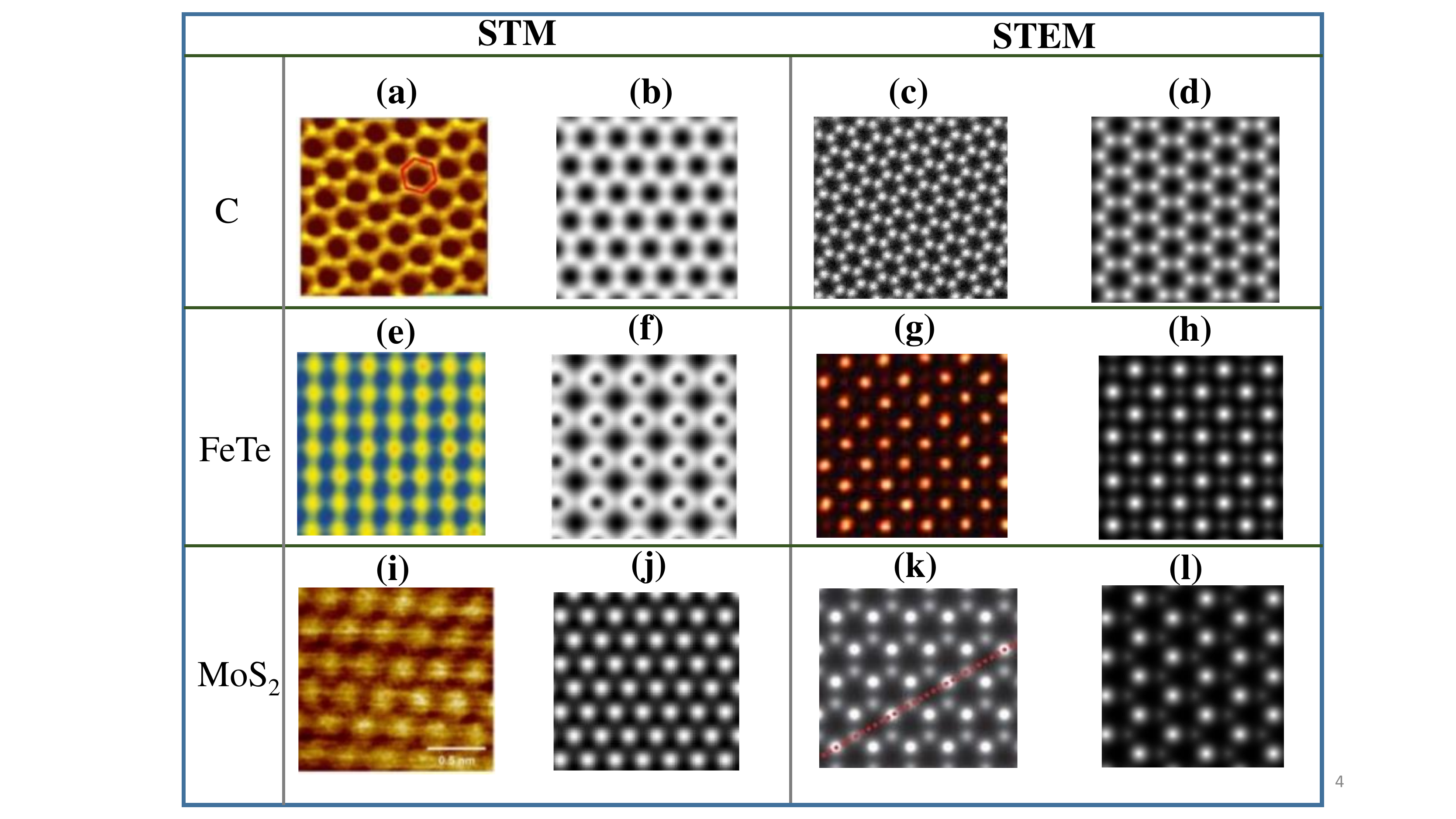}
    \caption{\label{fig:comp}\textcolor{black}{Comparison of our computational and previously reported experimental STM and STEM images for a few examples such as graphene, C (JVASP-667), FeTe (JVASP-6667), MoS$_2$ (JVASP-664). Fig. a,e,i shows experimental STM images \cite{li2009scanning,mills2016ripples,song2011direct} while b,f,j shows our simulated STM images. Similarly, c,g,k shows experimental \cite{o2022increasing,kang2020phase,migliato2020logic} and d,h,l shows simulated STEM images. Experimental images are reprinted with permission from American Physical Society and American Institute of Physics Publishing.}}
\end{figure}
We compare the STM and STEM simulated images of a few example 2D materials: graphene, FeTe and MoS$_2$, with experimental images in Fig.~\ref{fig:comp}. The left panel (panels a,e,i) show experimental STM images, while DFT based images are shown in the next column (in panels b,f,j). Similarly, we show the experimental STEM images for graphene, FeTe, and MoS$_2$ in Fig, c,g,k and corresponding convolution approximation based images are shown in Fig. d,h,l. Clearly, we find excellent qualitative agreement between the simulated and experimental images. Furthermore, we note that theoretical image datasets are larger and can be generated in a very controlled way compares to experimental images. Hence, a comparison for a few samples gives a qualitative idea that computer vision techniques applied to a theoretical image dataset should indicate similar confidence with respect to experiments.


\begin{figure}
    \centering
    \includegraphics[width=0.98\textwidth]{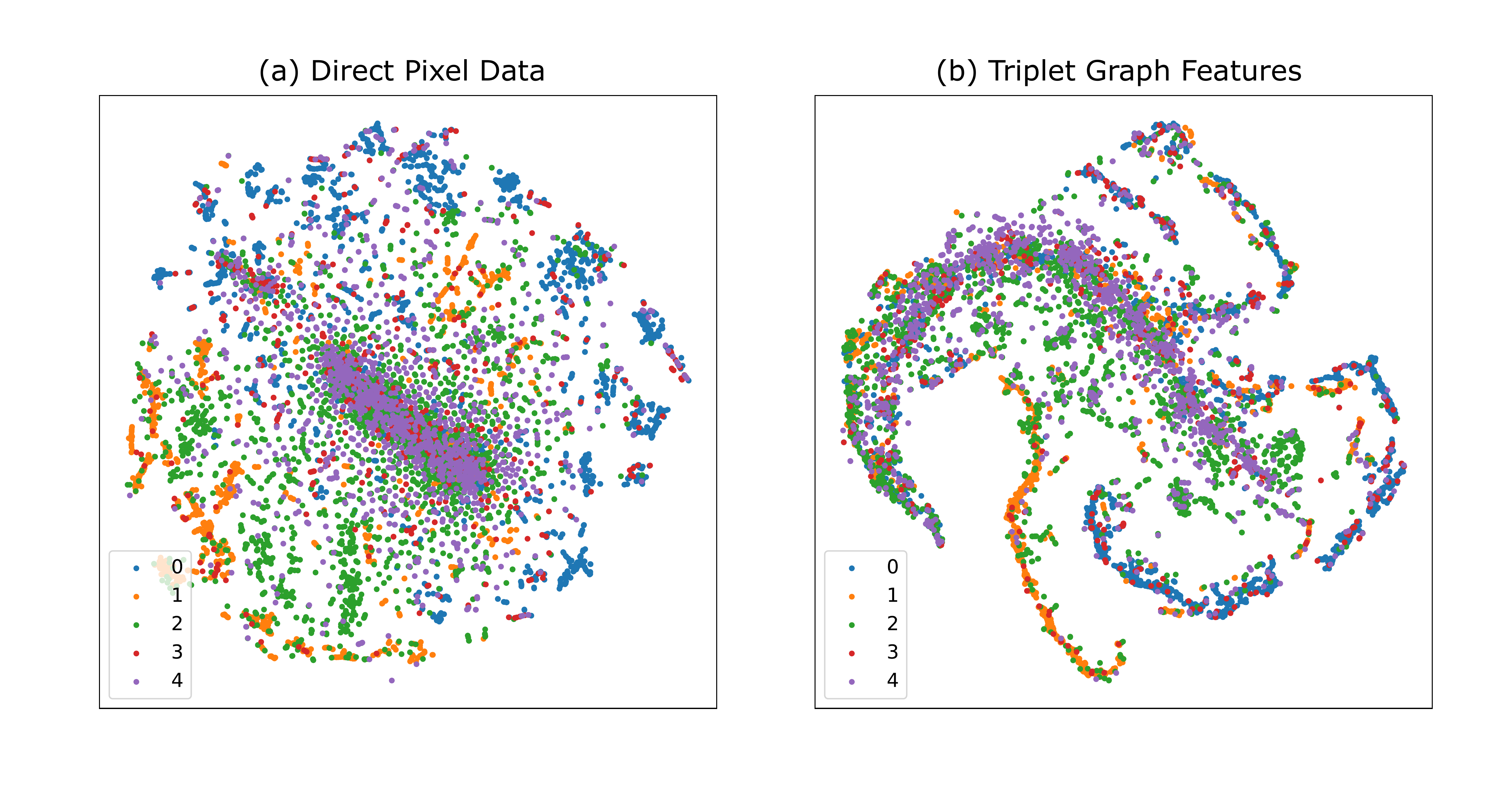}
    \caption{t-SNE visualization of the samples in the combined dataset including the JARVIS-DFT-2D and C2DB databases. In panel (a), the samples are featurized directly using the 256x256 pixel image intensity data. In contrast, in panel (b), the samples are featurized using their triplet (bond angle cosine) features in a graph construction. The distribution of triplet features are expressed as a 200-bin histogram ranging from -1 to 1. Here, we denote hexagonal, square, rectangle, rhombus and parallelogram classes as 0,1,2,3,4 respectively. We observe clusters of individual classes in both t-SNE figures.}
    \label{fig:tsne_comp}
\end{figure}

\subsection{Clustering analysis}
Next, we apply a machine learning clustering technique: t-SNE (t-distributed stochastic neighbor embedding) for visualizing the five types of 2D Bravais lattices in the JARVIS-DFT-2D STEM dataset. We first apply the t-SNE visualization directly to the simulated 256x256 pixel-based STEM image data for the 2D materials. These image array values for all the 2D materials in the datasets are converted into two-dimensions and are visualized in Fig.~\ref{fig:tsne_comp}. From this visualization, it is clear that hexagonal class 0), square (class 1), and rectangle (class 2) systems segregate into islands in the pixel space, while the rhombus (class 3) and parallelogram (class 4) systems show greater overlap with other classes, suggesting more likely mis-classifications. We then perform a similar visualization using the graph-based features for the STEM image dataset. Here, the list of triplet graph features for each image, representing bond angle cosines, are encoded in a 200-bin histogram before applying the t-SNE operation to investigate clustering. Once again, we find noticeable clusters of the hexagonal and square systems, but noticeable overlap between the hexagonal and rhombus classes as well as the rectangle and parallelogram classes. Such analysis provides a visualization of large dimensional data in a compact way to suggest that both image pixel values and graph features contain information about the Bravais lattices.

\subsection{U-Net and semantic segmentation}

\begin{figure}[hbt!]
    \centering
    \includegraphics[trim={0. 0cm 0 0cm},clip,width=0.98\textwidth]{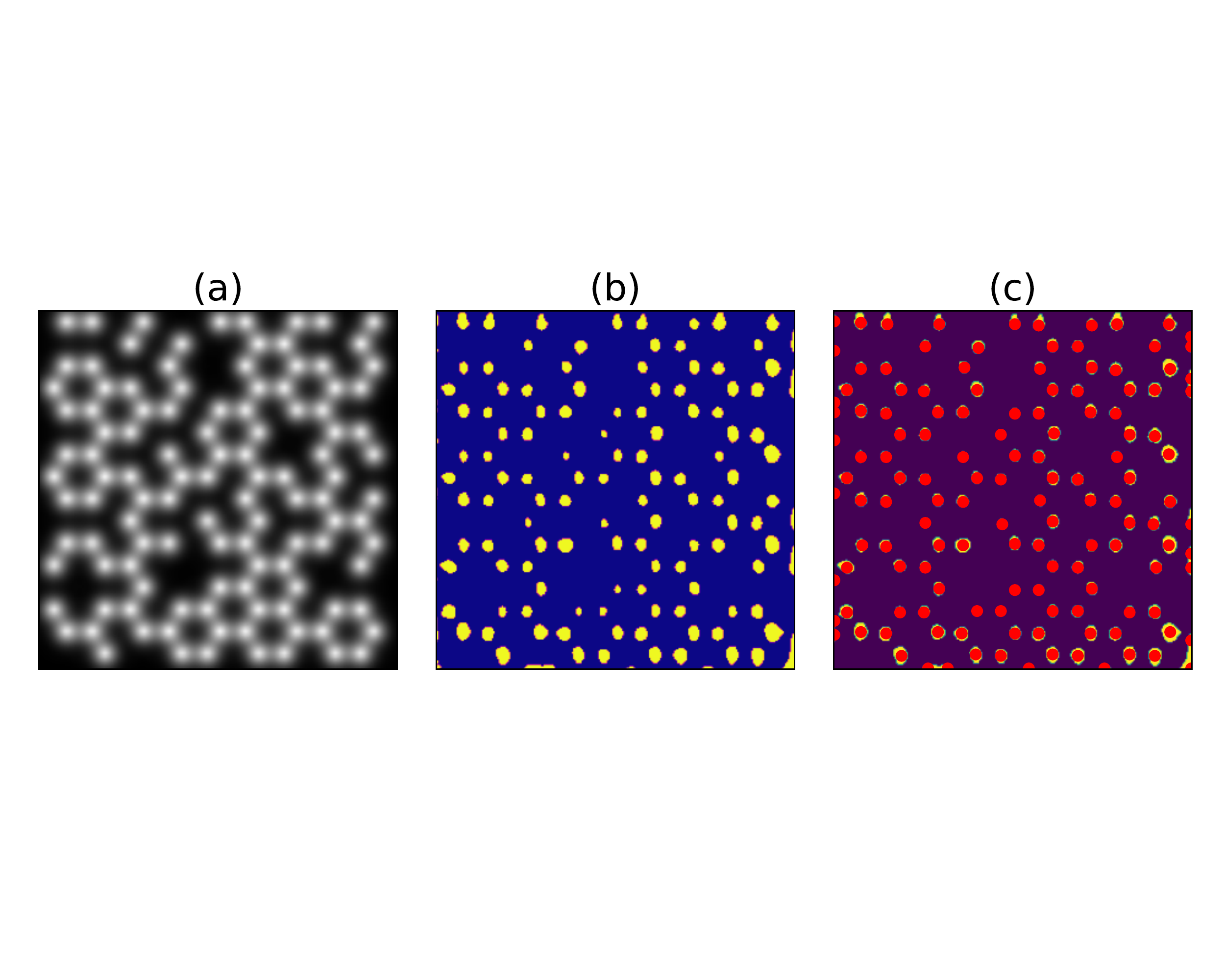}
    \caption{\label{fig:segment}\textcolor{black}{Application of semantic segmentation model to identify atoms and background from defective graphene system. In a) We generate synthetic graphene STEM images with vacancies and in b) use the semantic segmentation model. Finally, in c) we a blob detection algorithm to find number of atoms.}}
\end{figure}
Although, the clustering models are useful for identifying global features of images such as lattice system, more details can be obtained with pixelwise classification of images. We use U-Net to perform atom vs background classification task using the STEM datasets. U-net is a fully convolutional neural network with a U-shaped architecture consisting of a specific encoder-decoder scheme (as shown in Fig.~\ref{fig:archs}). The encoder reduces the spatial dimensions in every layer and increases the channels. On the other hand, the decoder increases the spatial dimensions while reducing the channels. We use the pretrained U-Net model in the SMP library and fine tune with the STEM dataset generated in this work for pixelwise classification of atom vs background classes. We find that high accuracy of 93.0 \% for the pixelwise classification of atom vs background task. As an application for this model, we first generate synthetic graphene STEM images with vacancies and use the semantic segmentation model on it as shown in Fig.~\ref{fig:segment}. We find find that the model can accurately identify atom and background from the image. Later, using scikit-image \cite{van2014scikit} based blob detection we can find the number of atoms in the image. With scikit-image, we can also get various statistics of the blobs such as the maximum, minimum, mean intensities in the blob etc.




\subsection{CNN Classifier}
\begin{figure}[hbt!]
    \centering
    \includegraphics[trim={0. 0cm 0 0cm},clip,width=0.98\textwidth]{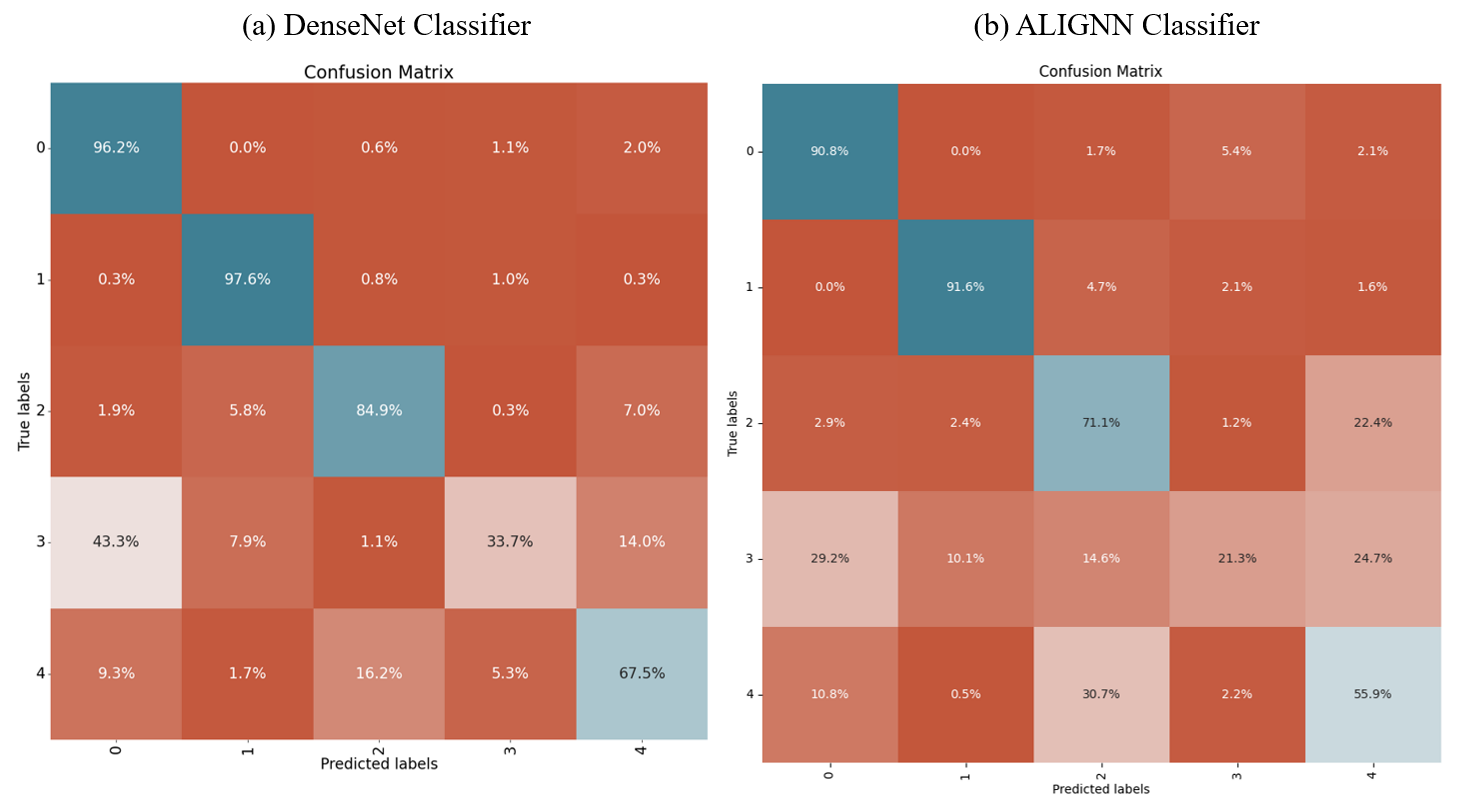}
    \caption{\label{fig:confmat}\textcolor{black}{Confusion matrix for classifying STEM images into five Bravais lattices using a) convolution neural network based Densenet, b) graph neural network based ALIGNN models. Here, we denote hexagonal, square, rectangle, rhombus and parallelogram classes as 0,1,2,3,4 respectively.}}
\end{figure}

Now, we use a supervised classification technique to classify images into the five 2D Bravais lattices: hexagonal, square, rectangular, rhombus (rectangular centered), and oblique (parallelogram)\cite{moeck2018accurate}. We use a well-known computer vision model: DenseNet to perform this task. In a DenseNet architecture, each layer is connected to every other layer, hence the name is given as Densely Connected Convolutional Network. For L layers, there are L(L+1)/2 direct connections. For each layer, the feature maps of all the preceding layers are used as inputs, and its own feature maps are used as input for each subsequent layers. DenseNet was developed to improve the vanishing gradient issues in deep convolutional neural networks. We fine-tune the DenseNet model \cite{huang2017densely} for 5 Bravais lattice classes using the STEM image dataset. We find that DenseNet provided an accuracy of 83.0 \%. We note that the baseline model for such classification task in 1/5 = 20 \%, so using machine vision techniques is clearly justified. In addition to overall accuracy of the model, confusion matrix of the classification task provide details for individual class performance as shown in Fig.~\ref{fig:confmat}a. Here, we denote hexagonal, square, rectangle, rhombus and parallelogram classes as 0,1,2,3,4 respectively.  We find that the trained model is highly accurate for hexagonal and square lattice but less accurate for rhombus and parallelogram classes, which can be attributed to less training data and higher complexity for these classes.

\subsection{ALIGNN-based GNN classifier}

\begin{figure}
    \centering
    \includegraphics{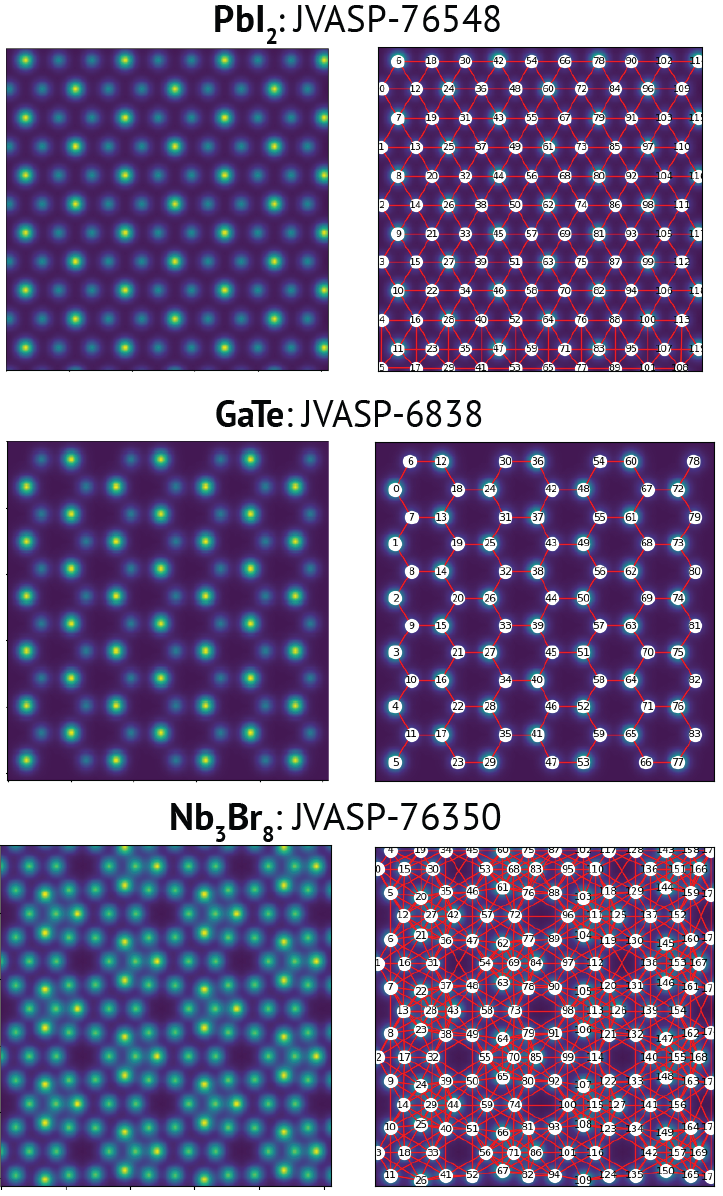}
    \caption{Computed STEM images (right) and overlaid graph construction (left) for three example materials in the hexagonal crystal system and (001) orientation. Although the lattices are visually diverse, the DenseNet and ALIGNN crystal system classifiers correctly categorize over 90 \% of the class 0 samples.}
    \label{fig:class0_examples}
\end{figure}

In the previous section, an image classifier was trained directly on the pixelated images. Here, we instead convert the image data into a non-Euclidean graph, which infers the connectivity of the objects in the image, allowing the usage of a graph neural network. Graph neural networks have been widely applied in the field of materials science as they allow for the structure of the material, along with composition-based features to be used in the prediction. Typically, graph neural networks are used to predict a material property from its structure, including both scalar quantities such as formation energy, bulk modulus, or band gap, or more recently, spectral quantities such as electron and phonon density-of-states or measured optical spectra (e.g. X-ray, infrared, Raman) \cite{choudhary2022recent}. To our knowledge, a graph neural network has not yet been applied to materials science image data, as it is in this work. The atomistic line graph neural network (ALIGNN) \cite{choudhary2021atomistic} was used, as it allows a hierarchy of structural features corresponding to single objects (i.e., atoms), pairs of objects (i.e., bond vectors), and pairs of bonds (i.e., bond angles). ALIGNN uses edge-gated graph convolution for updating nodes as well as edge features. One ALIGNN layer composes an edge-gated graph convolution on the bond graph with an edge-gated graph convolution on the line graph. The line graph convolution produces bond messages that are propagated to the atomistic graph, which further update the bond features in combination with atom features.

After the pixelwise classification, we can convert the data into graph that can capture non-Euclidean information of the images and can be used for advanced ML techniques such as the application of graph neural network. A example systems Pb$_2$I, GaTe and Nb$_3$Br$_8$ are shown in Fig. ~\ref{fig:class0_examples}. After the graph conversion we use the ALIGNN model for lattice classification and find a reasonable accuracy of  78 \%. Note that unlike the original ALIGNN model which uses atomic attributes such as electronegativity as node features, we use the blob-statistics (such as maximum, minimum and mean intensities in a blob) as the node attributes. Therefore, knowledge of chemistry is not required to train and execute the model. The confusion matrix for the model is shown in Fig.~\ref{fig:confmat}b. Although, the GNN-based models do not beat CNN models such as DenseNet, the framework for applying GNNs on images could be a powerful alternative tool for futuristic materials design because GNN-based methods can incorporate additional relationships and parameters that are not strictly related to the appearance of the image. Additionally, from Fig.~\ref{fig:confmat}, we find that both the CNN and GNN models work well for hexagonal and square lattice but they are less accurate for rhombus lattices.

\subsection{Autoencoders}

\begin{figure}[hbt!]
    \centering
    \includegraphics[trim={0. 0cm 0 0cm},clip,width=0.98\textwidth]{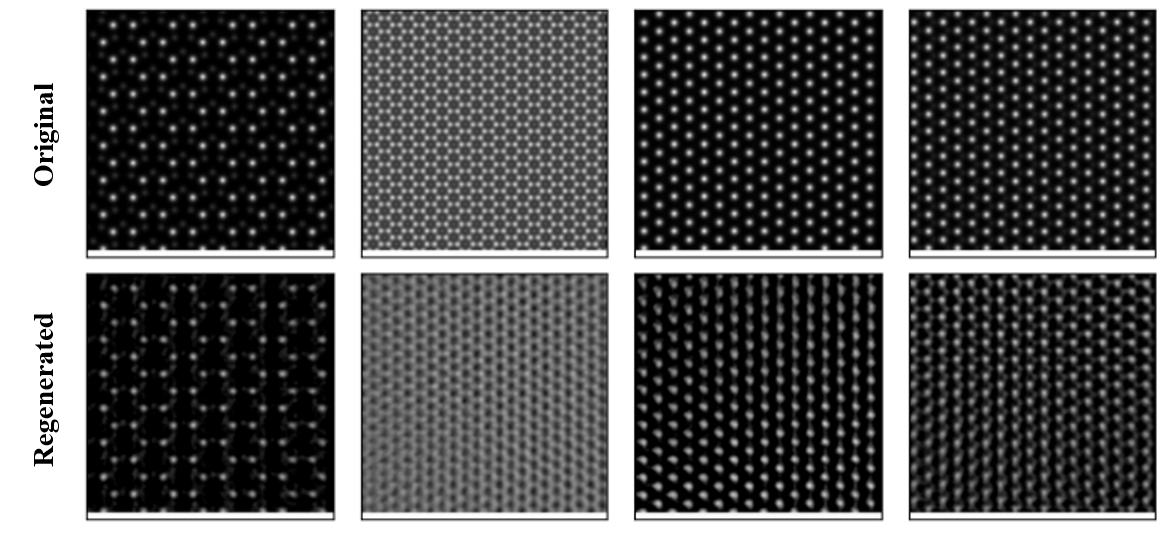}
    \caption{\label{fig:aencoder}\textcolor{black}{Original vs autoencoder regenerated images for a few random samples in the test dataset including Tl$_2$Se$_3$, Al$_2$O, ThF$_2$, and CuF. The top 4 images are the original STEM images while the lower ones are generated after autoencoder regeneration operation.}}
\end{figure}


Autoencoders (AE) are a special kind of neural network used to perform dimensionality reduction. We can think of autoencoders as being composed of two networks, an encoder and a decoder. The encoder learns a non-linear transformation from the original high-dimensional input space to a lower-dimensional latent space. A decoder learns a non-linear transformation that projects the latent vectors back into the original high-dimensional input space. We develop an auto-encoder to reduce the large pixel-dimension. Usually an image taken as 256x256 if flattened leads to 50176 (i.e., 256x256) dimension, which is quite high. According to  manifold hypothesis, the underlying structure of the data can be sufficiently described using only a few dimensions. Auto-encoders are well-known for such dimensional reduction tasks. We take the 256x256 image, and use an auto-encoder of dimension 1120. In Fig.~\ref{fig:aencoder}, we show the performance of the auto-encoders to reconstruct a few images in the test dataset. Clearly, the reconstructed images bears a lot of similarity to the original images suggesting that the autoencoders developed in this work can be used for AE-related tasks such as dimensionality reduction. 


\subsection{Super-resolution GAN}
\begin{figure}[hbt!]
    \centering
    \includegraphics[trim={0. 0cm 0 0cm},clip,width=0.98\textwidth]{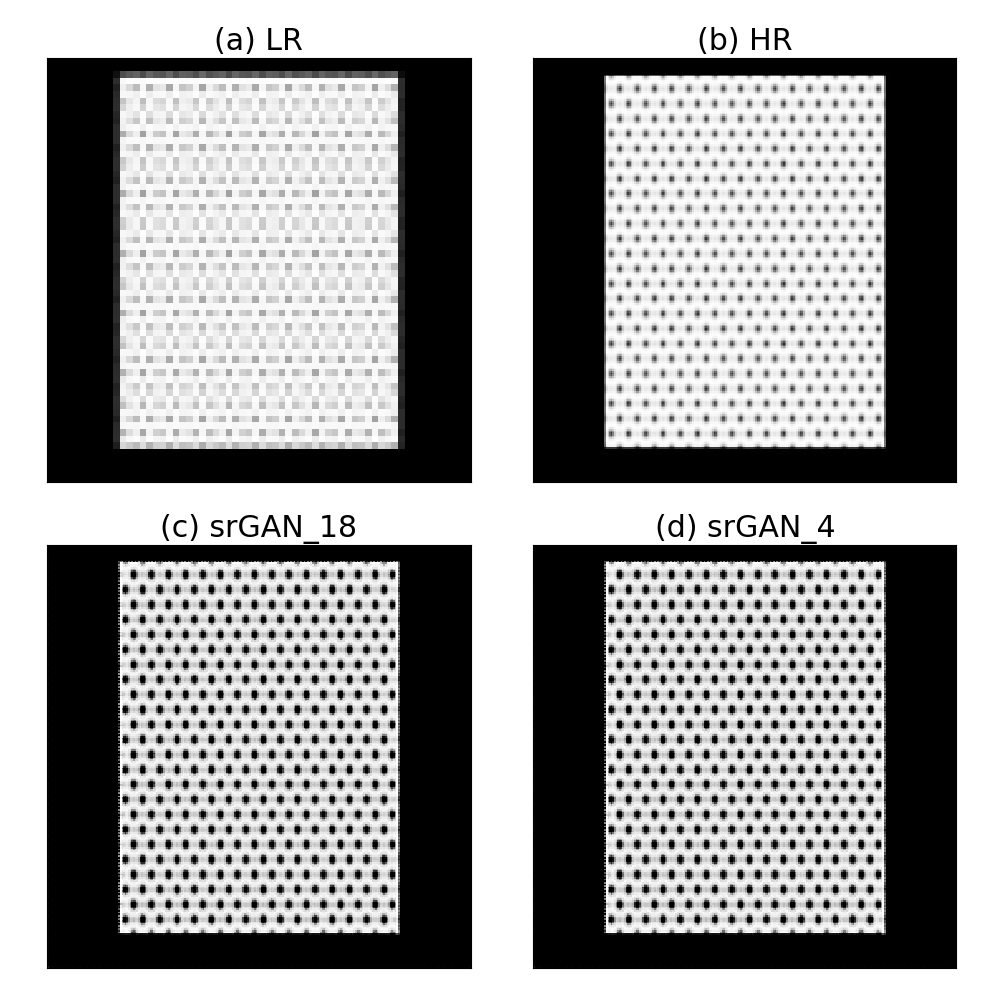}
    \caption{\label{fig:srgan}SR-GAN for enhancing image resolution. Examples for a) LR (low resolution), b) actual HR (high-resolution), c) SR-GAN VGG19's 18th layer prediction, d) SR-GAN VGG19's 4th layer prediction. We observe shallower layers of VGG19 gives similar predictions in terms of resolution enhancement as that of deeper layers.}
\end{figure}

Now, we use generative adversarial network (GAN) for enhancing the resolution of STEM images. We generate a dataset of 4-times low resolution (i.e., 64x64 instead of 256x256 size images) and using Super-Resolution Generative Adversarial Networks (SRGAN) \cite{ledig2017photo} architecture, we develop a model to enhance resolution. We find a generator loss of 0.306 and discriminator loss of 0.458. Using this model, we generate a high-resolution image in the test set and show the results in Fig.\ref{fig:srgan}. The low-resolution image is shown as LR while the high-resolution image is shown as HR. SRGAN uses VGG19 (visual geometry group  convolutional neural network that is 19 layers deep) \cite{simonyan2014very} as feature extractor, however we notice that using shallow VGG19 feature extractor such as 4th layer (Fig. c) is equivalent to deeper layers such as  Fig. d. Hence, a lower level feature extractor in VGG19 should be enough for resolution enhancement purposes.

\subsection{Extracting images from arXiv dataset}

\begin{figure}[hbt!]
    \centering
    \includegraphics[trim={0. 0cm 0 0cm},clip,width=0.98\textwidth]{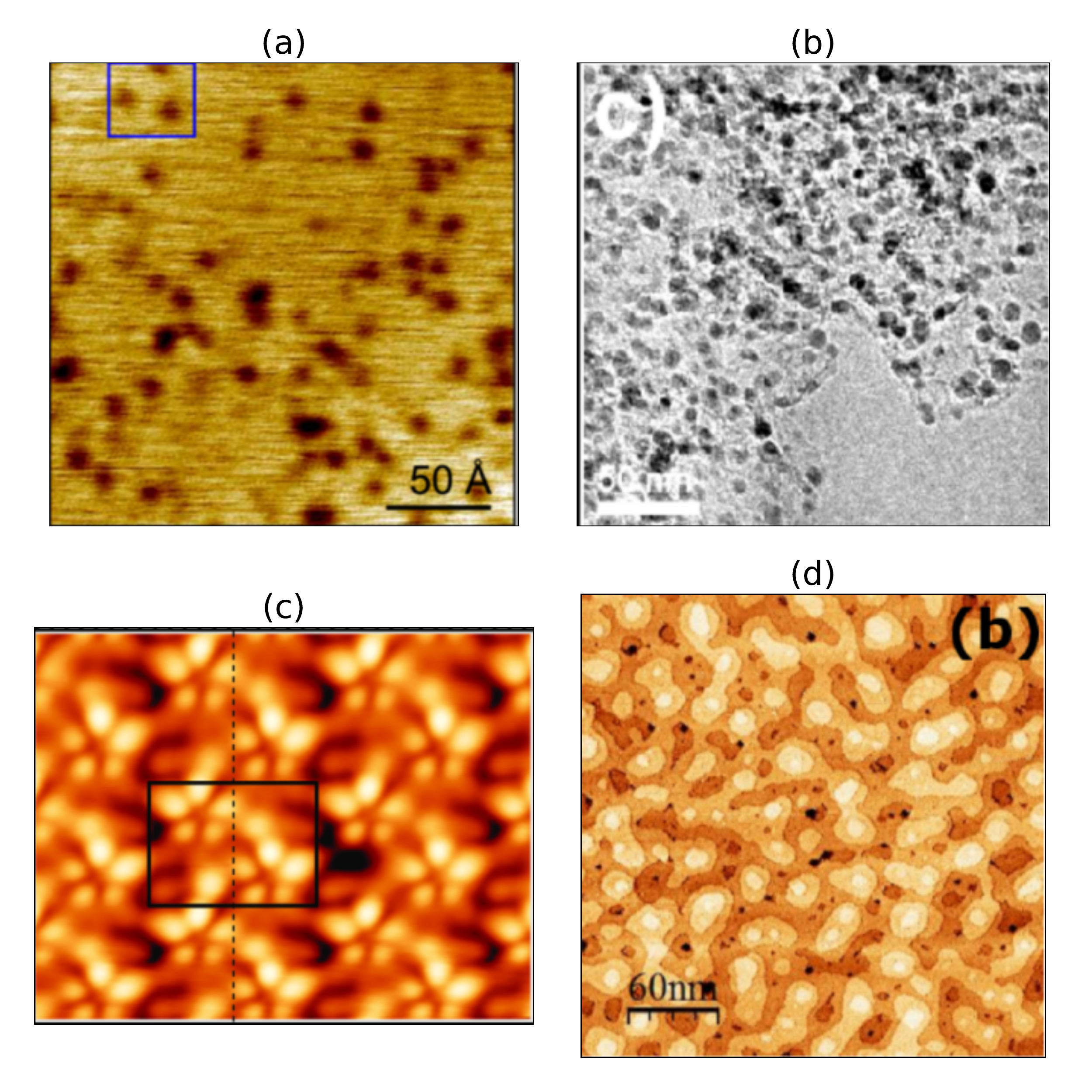}
    \caption{\label{fig:arxiv}\textcolor{black}{Example micrographs retrieved from the arXiv dataset. These figures are taken from: a) arXiv:0708.2306 for Dy \cite{wegner2007effect}, b) arXiv:0805.3416 for FeC$_{32}$H$_{16}$N$_8$ \cite{klinke2007iron} c) arXiv:0807.3875 for Si \cite{battaglia2008new} d) arXiv:0902.0626 for La$_{0.67}$Ca$_{0.33}$MnO$_3$ \cite{kelly2009correlations}.}}
\end{figure}

Next, we use natural language processing with the open access arXiv dataset to curate datasets of STM and STEM images from the available literature. The arXiv dataset was previously used for ChemNLP project and has chemistry based information for the systems as well. We found more than 500 STM and 1500 STEM images from a simple search of STM, STEM in abstracts for the condensed-matter physics articles in arXiv. We further searched for such entries in the figure captions of the dataset and found more than 1000 such images that can be useful for image analytics. We show a few of the images obtained from the arXiv dataset in Fig.\ref{fig:arxiv}. We provide the list of links to the images and corresponding papers in the AtomVision library.

\subsection{Experimental image dataset}

\begin{figure}[hbt!]
    \centering
    \includegraphics[trim={0. 0cm 0 0cm},clip,width=0.98\textwidth]{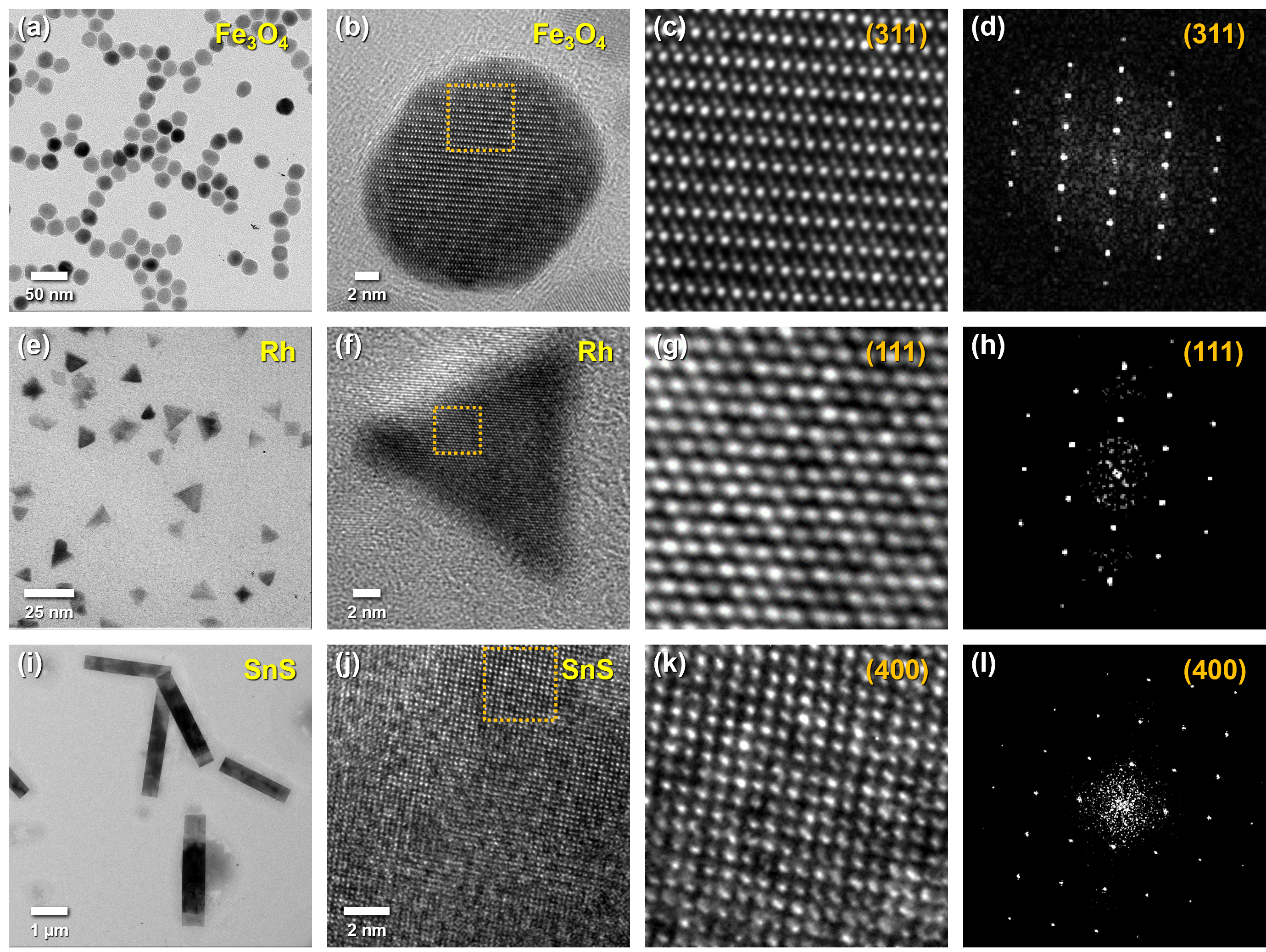}
    \caption{\label{fig:tem}\textcolor{black}{TEM and corresponding HRTEM images of solution-synthesized nanostructures.  (a,b) Fe$_3$O$_4$ spherical nanoparticles, (e,f) Rh triangular nanoplates, and (i,j) SnS micron-sized nanoribbons.  Regions of the HRTEM image indicated by dashed lines were magnified and analyzed by fast Fourier transform revealing crystallographic assignments of (c,d) Fe$_3$O$_4$ (311), (g,h) Rh (111), and $\alpha$-SnS (400).}}
\end{figure}




In addition to computational and NLP based microscopy image datasets, we also develop our own experimental image dataset in AtomVision. AtomVision provides a flexible and easy to use metadata capture template that can capture experimental set up across various microscopy instruments. This is motivated by previous works in this field such as refs. \cite{nguyen20174ceed,taillon2021nexuslims}. Such infrastructures provide frontend and backend meta-data capture schema and frameworks to capture and curate microscopy image data. Having an integrated framework for both experimental, computational and image based datasets in AtomVision will allow the investigation of several important challenges such as reproducibility, ground truth data and uncertainty in measurements. We show a few examples of TEM image dataset of nanoparticles in Fig.\ref{fig:tem}. These nanoparticles TEM images are labelled by their different facet orientations leading to a huge variability in the dataset for the same material. Currently, the experimental image dataset contain a few hundreds of TEM images for various materials and their facets including Pt, Pd, Rh, Au, Fe$_2$O$_3$, SnS etc. and we plan to continuously grow the dataset. We can integrate the AtomVision framework with NIST's and other micorscopy measurement labs in the future to leverage several tools and datasets available in the library.

In summary, we have developed an integrated and general-purpose machine-vision library especially for atomistic images. The dataset in AtomVision consists of both computational, experimental and literature based images providing a wide variety for general applications including machine/deep learning applications. The dataset mainly consists of  scanning tunneling microscopy (STM) and scanning transmission electron microscopy (STEM) images and the framework would allow other atomic image datasets as well. There are numerous image machine learning techniques and we demonstrated applications of few of them  including convolution neural network,  graph neural network, fully convolution neural network, generative adversarial network etc. Especially, the application of graph neural network such as ALIGNN on atomistic images provide a new paradigm for atomistic image analysis. The well-curated image dataset from experiments as well as the computational images can serve as reference for many scientific applications.






\section{Data availability statement}
The data that support the findings of this study are openly available at the following URL: \url{https://github.com/usnistgov/atomvision}.

\section{Acknowledgments}
We thank the National Institute of Standards and Technology for funding, computational, and data-management resources. K.C. thanks the computational support from Extreme Science and Engineering Discovery Environment (XSEDE) computational resources under allocation number TG-DMR 190095. Contributions from K.C. were supported by the financial assistance award 70NANB19H117 from the U.S. Department of Commerce, National Institute of Standards and Technology. 


\bibliographystyle{unsrt}
\bibliography{cross_ref}

\end{document}